\documentclass[12pt]{article}
\usepackage{graphicx,amsmath,amssymb,euscript}

\newcounter{aaa}
\newenvironment{teor}[2][{}]{\begin{trivlist}\refstepcounter{aaa}%
\labelsep=0pt\item[\bfseries \theaaa. #2. ]#1}%
{\end{trivlist}}
\newcommand{\ssy}[5]{#1,  {\it #2}  {\bf #3} #5 (#4)\rlap{.}}

\DeclareSymbolFont{forvarg}     {U}{txmia}{m}{it}
\DeclareMathSymbol{\phiup}{\mathord}{forvarg}{"1E}
\DeclareMathSymbol{\varg}{\mathord}{forvarg}{"31}
\DeclareMathAlphabet{\mathpzc}{OT1}{pzc}{m}{it}
\newcommand{\CauD}{\mathpzc{D}}

\newcommand*{\ogr}[2]{#1\, \rule[-0.76em]{0.4pt}{1.4em} \,
\raisebox{-.55em}{$\displaystyle{}_{#2}$}  }%
\DeclareMathOperator{\Bd}{Bd}

\DeclareMathOperator{\Vn}{Int}

\title{The speed of gravity in general relativity}

   \author{S. Krasnikov}

\begin{document}
\maketitle
\begin{abstract}
    The question is discussed of what is the speed of gravity (at the
    fundamental non-perturbative level). The question is important, if
    nowhere else, in
    discussing the problem of information ``lost" in black
    holes. It turns out that the duly defined
    ``gravitational signal" generally may be causal, superluminal and
    ``semi-superluminal". In the class of globally hyperbolic spacetimes the
    two last varieties coincide. And if some (often imposed, but not always
    satisfied) conditions hold, the signals may be \emph{only} causal. In
    this sense the speed of gravity does not exceed the speed of light.
\end{abstract}

\section{Introduction}
The question of how large is the speed of gravity may seem meaningless at
first sight. In general relativity there is no ``gravitational field", which
could ``propagate". There is only a (curved) four-dimensional spacetime
and the word ``gravitational"  essentially means ``related to the shape of
that spacetime". And one does not ask: ``What is the speed of a sphere's
being a sphere?"

Imagine, on the other hand, an observer  $ \mathcal{O}$ who watches a
rock held by a person   $ \mathcal{P}$.   $ \mathcal{O}$ wants to know
whether the rock will be thrown. Apparently, this event |   throwing of the
rock, let us denote it $s$ | can be detected by   $ \mathcal{O}$ even
without looking at $\mathcal P$. Indeed, the moment the rock starts to
move, its distance from the observer will start to change. So, if  the
observer's apparatus is good enough, it will detect ---  at some moment $q$
--- changes in the gravitational field of the rock, which will mean  that the
rock is thrown. It seems absolutely natural to interpret the event  $q$ as
receiving by the observer a \emph{gravitational signal} from  $
\mathcal{P}$. And it is equally natural to ask: How soon after $s$ can $q$
happen? Can, in particular, $q$ happen \emph{before} the first photon
from $s$ will reach the observer. In other words, what is the speed of
gravity and are there any restrictions on this speed?

Problems in answering those questions arise immediately as one tries to
make the outlined naive picture more precise. Indeed, we have described
$q$ as the point at which the observer notices the first changes in the
metric. But changes \emph{with respect to what}? It is definitely
\emph{not} changes w.~r.~t. some previous moment of time (in a
non-static world the metric changes all the time even while the rock is at
rest). Neither it is a deviation from some background (contrary to what is
often assumed in the gravitational waves theory), because generally there
is no way to unambiguously split the metric into ``background" and
``perturbation" (cf.~\S 35.1 in \cite{MTW}). Actually, what is meant are
changes w.~r.~t. what \emph{would} there be there in the world in which
the rock is not thrown. So, we have to compare \emph{two different
spacetimes}: the spacetime, denote it $M_1$, in which  $\mathcal P$
throws the rock in $s$ and the spacetime $M_2$ in which he does not.
Proceeding along these lines one has to solve the following two problems
(the former being essentially a matter of language in which the latter is to
be discussed):
\begin{enumerate}
  \item To tell whether the geometry about $q_1\in M_1$ differs from
      what would have been there were the rock not thrown, we must find
      a point  $q_2\in M_2$ corresponding to $q_1$ and compare
      somehow the metric in $q_1$ with that in $q_2$. But no canonical
      ways are seen how to bring points of non-isometric spacetimes in
      one-to-one correspondence.
  \item Our goal is to establish  (an analog of) the cause--effect
      relation\footnote{Note that ``causally related" in this context
      \emph{cannot}, of course, mean ``connected by a non-spacelike
      curve"; the contrary would mean that we postulate the impossibility
      of superluminal gravitational signals \emph{a priori}.}. The
      introduction of the pair $(M_1,M_2)$ instead of a single spacetime
      enables us to formulate the task in terms of   ``changes in the
      geometry": the problem now reduces to telling the changes caused by
      $s$ from any other possible differences between $M_1$ and $M_2$
      (such as caused, for instance, by a set of independent events). This,
      however, cannot be done (exclusively) on the basis of equations of
      motion. The point is that the concepts --- such as  the field value , the
      energy density, the group velocity, etc. --- pertinent to   evolution of a
      field, have no ``self-evident" relation\footnote{As is verified by
      unsuccessful efforts to identify the speed of signal propagation with
      the phase or group velocities.} to concepts
      --- such as signal, cause, etc. --- determining which elements of the
      theory are considered freely specifiable. To relate these two parts of a
      theory one need a convention (postulate, definition) additional to the
      equation of motion.
\end{enumerate}

\section{Alternatives}

We start with formalizing the idea that two different spacetimes may be
``same up to some event $s$'', \emph{i.~e.,} may have a ``common origin".
\begin{teor}{Definition}\label{def:alt}
A pair of pointed inextendible spacetimes $(M_k,\varg_k,s_k),\  k=1,2$ is
called an \emph{alternative}, if   there exists a connected open past set
$N_1\supset I^-(s_1)$ and an isometry $\phiup\colon\ N_1\to M_2$
such that $\phiup(I^-(s_1))=I^-(s_2)$.
\end{teor}
Thus   $M_1$ and $M_2$ are different extensions of the same (extendible) spacetime $N_1$.

 To overcome the first of the problems indicated in the Introduction we concentrate on the
points which are \emph{not} affected by $s$ (the ``common origin" of the
two universes). For a given alternative the pair $N_1,\phiup$
need not be unique, there may exist a whole family
$\{N^\alpha_1,\phiup^\alpha_1\}$ of such pairs. By $N^*_k$ and
$\phiup^*$ we shall denote the \emph{maximal} elements of this family,
i.~e., such that
\[
N^\alpha_1 \subset N^*_1,\quad \phiup^\alpha=
\phiup^*\ogr{}{N^\alpha_1},
\qquad\forall\,\alpha.
\]
It is  $N^*_1$ (and   $N^*_2$ isometric to it) that describe the
above-mentioned common origin.

\begin{teor}{Remark}%\label{}
The reason to require $N^*_k$ to be past sets is simple: even if somewhere
in $M_k$ there are isometric domains, they can hardly be reckoned among
those constituting the common origin of $M_k$, if their inhabitants
remember different histories.
\end{teor}

A more local characteristic of gravitational communication is the front | an
analogue of the signal.
\begin{teor}{Definition}
The sets
 $\EuScript N_k\equiv \Bd  N_k^*,\  k=1,2$ will be  termed
 \emph{fronts}. A front $\EuScript
 N_k$ is called \emph{superluminal} if $\EuScript N_k \not\subset
 \overline{J^+(s_k)}$.
\end{teor}
\section{The speed of gravity in general relativity}
Now given a theory (by which I mean a set of  matter fields with equations
relating them to geometry) one makes the following step and adopts a
convention defining what in that theory is considered freely specifiable.
Namely, one decides what class of alternatives are \emph{admissible},
i.~e., in what cases the difference between the spacetimes $(M_1,\varg_1)$
and $(M_2,\varg_2)$ is attributable to the event $s$ (modeled in our
approach by two points at once: $s_1\in M_1$ and  $s_2\in M_2$). This
being done, one can check whether the fronts of admissible alternatives
may be superluminal and, if so, decides whether the theory should be
dismissed on that ground.

Let us apply the abovesaid to general relativity. To this end we, first,
assume that the Universe is described by a spacetime on which some
``matter fields" are defined. The latter are subject to some conditions, see
below, and the metric solves the Einstein equations. As was argued in the
Introduction, however, to specify the theory we must adopt one more
convention.
\begin{teor}{Convention}\label{conv:admis}
    An alternative
    $(M_1,M_2)$, where both spacetimes are globally hyperbolic, is
    admissible (\emph{i.~e.,}      $M_1$ and
    $M_2$ are regarded as differed by $s$ and its
    consequences, rather than two \emph{initially} different
    universes) only if in
    $M_1$ and $M_2$ there are  Cauchy surfaces $\EuScript S_1$ and
    $\EuScript S_2$, respectively, such that  $\EuScript S_2- s_2 =
    \phiup^*(\EuScript S_1 - s_1)$ and the values of all fields and their
    derivatives in the corresponding points of $\EuScript S_1 - s_1$ and
    $\EuScript S_2 - s_2$ are same\footnote{It would be more consistent
    to speak of infinitely small neighborhoods of $s_{1,2}$ rather than
    of the points \emph{per se}, but for simplicity we shall neglect such
    subtleties.}.
\end{teor}
It should be stressed that this convention is an independent element of
theory. Even though it seems so ``natural"  (and is often regarded  as
``self-evident", see \cite{Low} for example), it does not, in fact,
     follow from, say, the field equations, or any first principles. Suppose
     for instance, that each time one makes the metric  strictly flat on
     some three-dimensional disk $D_1$ (and \emph{only} in such a case)
     it turns out that the metric is also strictly flat on a $D_2$ lying on the
     same Cauchy surface $\EuScript S$ (say, a kilometer to the
     north-west of $D_1$). Be such a phenomenon discovered, one
     probably would wish to call it superluminal signalling.
     Convention~\ref{conv:admis} would have to be abandoned [because
     no admissible alternative to $(M,\varg,s)$ is ever observed, where
     $M\supset D_1\supset s$ and $g$ is a metric flat on $D_1$], even
     though the  field equations and the Einstein equation are not
     compromised.

Now suppose,  that both  $M_1$ and $M_2$ are globally hyperbolic and
the alternative $(M_1,M_2)$ is admissible. The question of whether the
$k$-th front is superluminal reduces to the question of whether $\phiup$
(or, correspondingly,  $\phiup^{-1}$) can be extended to the entire $M_k-
J^+(s_k)$. But the global hyperbolicity of $M_1$, $M_2$ implies the
equality
\[
M_k- J^+(s_k) = J^-(\EuScript S_1-s_1)\cup\,\Vn\CauD (\EuScript S_k-s_k),
\]
where $\CauD (X)$ denotes the Cauchy development of $X$. By the
theorems of existence and uniqueness for the Einstein equations, the
equality of the data on a surface implies (under some assumptions which
we shall discuss in a moment)  isometry of the resulting Cauchy
developments. So, the domains $\Vn\CauD (\EuScript S_k-s_k)$ are
isometric and we conclude that $N_k^*$ do include the whole $M_k-
J^+(s_k)$. Thus, neither of the fronts is superluminal. In this sense general
relativity forbids superluminal signalling: \emph{the speed of gravity
does not exceed the speed of light}.

The mentioned uniqueness theorems  for  solutions of the Einstein
equations are proved under some ``physically reasonable" assumptions on
the properties of their right hand sides. In particular, three such
assumptions are formulated in \cite{HawEl}. The first of them is
essentially the condition that   the | duly defined | speed of  ``matter
signals" does not exceed the speed of light, the second one is a stability
requirement, and the third one restricts the stress-energy tensors to
expressions polynomial in $g^{ab}$ (the corresponding restriction in
\cite{Wald_GR} allows also first-order derivatives of the metric). This last
assumption is \emph{known} to fail in many physically interesting
situations. In particular, vacuum polarization \emph{typically} leads to
appearance of terms with second-order derivatives of the metric (like the
Ricci tensor) in the right hand side of the Einstein equations, see
\cite{BirDav}. So, one can expect that the gravitational signals may
propagate faster than light on the horizon of a black hole, where
semi-classical effects are strong.
\section{Semi-superluminal signals}
An important dissimilarity between the concepts of the matter field signal
and the front is that a single event is associated with \emph{one} signal,
but \emph{two} fronts and the latter do not have to be superluminal both
\emph{at once}.
\begin{teor}{Definition} An alternative is called \emph{superluminal}, if
both fronts are superluminal and \emph{semi-superluminal} if so is only
one.
\end{teor}
Consider a world  $M_1$ where a photon (or some other \emph{test}
particle) is sent from the Earth (the event $s_1$) to arrive  to a distant star
at some moment $\tau_1$ ($\tau$ parametrizes the world line of the star).
Let, further, $M_2$ be the world which was initially the same as $M_1$,
but in which a huge spaceship is sent to the star instead of the photon (the
start of the spaceship is $s_2$). On its way to the star the spaceship warps
and tears the spacetime by exploding passing stars, merging binary black
holes and otherwise employing immense energies and matter with little
understood properties. If the Causality principle (by which the assertion is
understood that the speed of matter field signals does not exceed the speed
of light) holds in $M_2$, the spaceship arrives at the star later than the
photon emitted in $s_2$,  but nevertheless it is imaginable that its arrival
time $\tau_2$ is \emph{less} than $\tau_1$ (so, the speed of light in one
world does not restrict the speed of the spaceship in the other).  Indeed, the
assumed prohibition of superluminal signalling  in $M_1$  does not
prevent $\EuScript N_1$ from being spacelike, because $\EuScript N_1$
does not correspond to any matter field signal  (the world line of the
spaceship does \emph{not} belong to $\EuScript N_1$). The pair
$(M_1,M_2)$ is an example of what we have called the semi-superluminal
alternative. In a theory admitting the alternatives of this kind
faster-than-light trips do not require violations of the Causality principle.
\begin{teor}{Example}\label{ex:hyperj}
Let $M_1$ be the Minkowski plane and $s_1$ be the point with the
coordinates $x=-1$, $t=-3/2$. Let, further,  $M_2$ be the spacetime
obtained by removing the segments $t\in[-1,1]$, $x=\pm 1$, from the
Minkowski plane, see the figure, and gluing the left/right bank of either
cut to the right/left bank of the other one.
\begin{figure}[tbh]\begin{center}
\includegraphics[width=\textwidth]{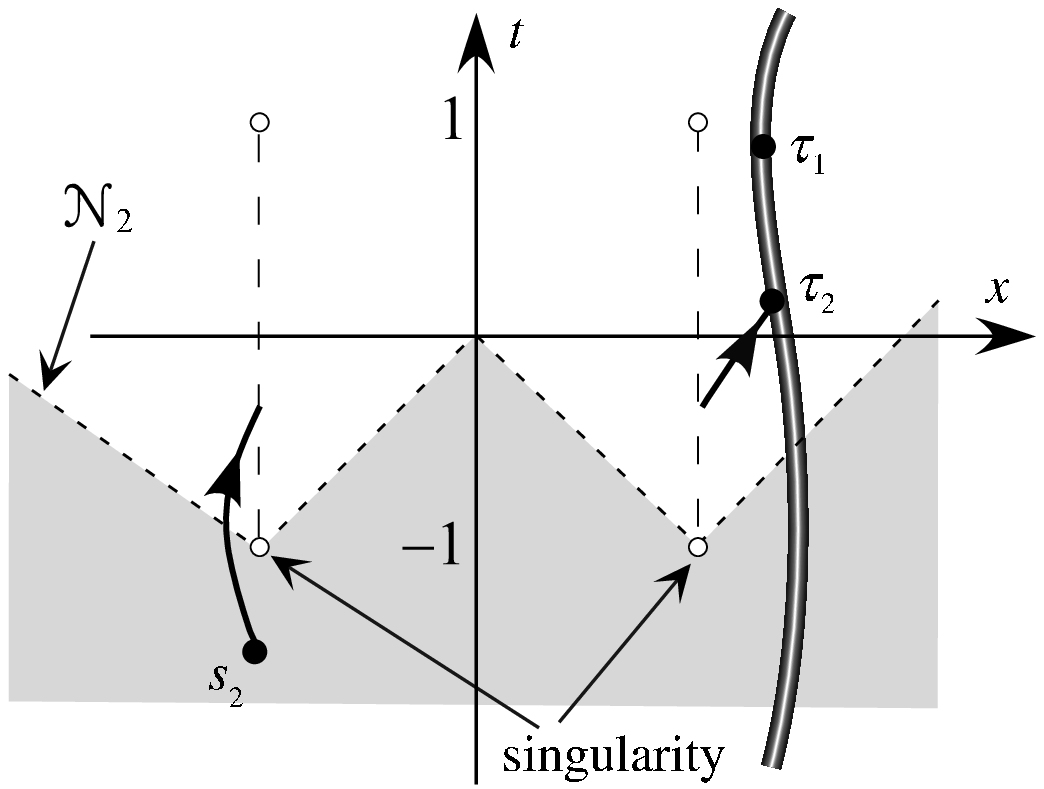}
\end{center}\caption{The spacetime $M_2$. The arrowed thick  line depicts\label{fig:hyperj} a
continuous timelike curve. The dashed lines show where the banks of the
former cuts are glued together, the gray region is  $ N^*_2$.}\end{figure}
 Then $N^*_1$ is the complement to the union of
two future cones with the vertices at the points $t=\pm x=-1$. That
$N^*_1$ is maximal indeed is clear from the fact that otherwise being a
past set it would contain a past directed timelike curve $\lambda$
terminating at one of the vertices, while $\phiup(\lambda)$ does not have
a past end point due to the singularity.

Obviously $\EuScript N_1\not\subset \overline{J^+_{M_1}(s_1)}$ in this
case, so the front $\EuScript N_1$ is superluminal. However, in the
(empty) spacetime $M_1$ the surface $\EuScript N_1$ does not
correspond to any matter field signal, so its superluminal character does
not violate the Causality principle. At the same time $\EuScript N_2$ is
not superluminal (and so, the alternative $(M_1,M_2)$ is
semi-superluminal). Even though the spaceship arrives to its destination
sooner than the photon in $M_1$ does ($\tau_2<\tau_1$) it does not
outrun the photons in $M_2$, so the Causality principle holds there too.
\end{teor}

The spacetime $M_2$ is singular (though the singularity is absolutely
mild). So, one may wonder if the whole effect is related to this fact. Indeed,
the assumption that both spacetimes in an alternative are globally
hyperbolic makes the prohibition of superluminal and semi-superluminal
alternatives equally strong.
\begin{teor}{Proposition}\label{prop:hot1}
A semi-superluminal alternative $(M_1,M_2)$ is superluminal if both
$M_1$ and $M_2$ are  globally hyperbolic.
\end{teor}
(The proof is rather technical and will be published elsewhere).

\section*{Acknowledgements}
 I wish to thank R. Zapatrin for helpful
discussions on the subject.

\end{document}